\documentstyle[eqsecnum,aps]{revtex}
\begin{document}
\title{Wave equations for the perturbations of a charged black hole}
\author{Zolt{\'{a}}n Perj{\'{e}}s}
\address{KFKI Research Institute for Particle and Nuclear Physics,\\
H--1525, Budapest 114, P.O.B.\ 49, Hungary\\
{\tt e-mail: perjes@rmki.kfki.hu}ÿ}
\date{\today }
\maketitle

\begin{abstract}
A pair of simple wave equations is presented for the symmetric gravitational
and electromagnetic perturbations of a charged black hole. One of the
equations is uncoupled, and the other has a source term given by the
solution of the first equation. The derivation is presented in full detail
for either axially symmetric or stationary perturbations, and is quite
straightforward. This result is expected to have important applications in
astrophysical models.

PACS Numbers: 04.70.-s, 97.60.Lf
\end{abstract}

\section{\bf Introduction}

Recent discoveries\cite{XMM} of astrophysical X-ray sources of extraordinary
intensity have instigated research on the nature of these objects. Current
speculation involving mechanisms such as the Blandford-Znajek effect\cite
{Znajek} in the vicinity of black holes has been hampered by a lack of a
viable theory of perturbations of charged black holes.

In the three decades following the discovery of the master equation of the
perturbative treatment of gravitational, electromagnetic and Weyl neutrino
fields in the presence of an uncharged black hole\cite{Teukolsky}, much
effort has been spent\cite{Fackerell,Chandrasekhar} on finding a
corresponding description of the fields for charged Kerr-Newman black holes.
It has been shown\cite{Crossman}, for example, that a decoupled equation
exists for either the electromagnetic or the gravitational component.

The purpose of this paper is to show for the first time that master
equations do exist for {\it both the electromagnetic and the gravitational
perturbation components} of the charged black hole. Symmetry will be assumed
to hold in order to keep the discussion simple. The required symmetry can be
any one-parameter group of isometries with a time- or spacelike Killing
vector.

It is hoped that the perturbation formalism presented in this paper will
find important applications in creating models of relativistic sources of
electromagnetic and gravitational radiation, both in analytical and in
numerical approaches. Axisymmetric perturbations, characterized by a
vanishing angular frequency of the normal modes, for instance, may already
suffice to construct models of astrophysical X-ray sources.

In the next section, we shall briefly review the relevant theory.

\section{\bf Einstein-Maxwell fields with a symmetry}

An Einstein-Maxwell system with one Killing vector, spacelike or timelike,
may be fully characterized by the complex 3-covectors \cite{Perjes}: 
\begin{equation}
{\bf G}=\frac{\nabla {\cal E}+2{\bar{\Phi}}\nabla \Phi }{2({\rm Re}{\cal E}+{%
\bar{\Phi}}\Phi )},\qquad {\bf H}=\frac{\nabla \Phi }{({\rm Re}{\cal E}+{%
\bar{\Phi}}\Phi )^{1/2}}
\end{equation}
where ${\cal E}$ and $\Phi $ are the complex gravitational and
electromagnetic Ernst poten\-tials\cite{Ernst}, respectively. In the
notation referring to the metric of the three-space of Killing trajectories,
the field equations can be written 
\begin{equation}
R_{\mu \nu }=-G_\mu {\bar{G}}_\nu -{\bar{G}}_\mu G_\nu +H_\mu {\bar{H}}_\nu +%
{\bar{H}}_\mu H_\nu  \label{eq:gheq0}
\end{equation}
\begin{eqnarray}
(\nabla -{\bf G})\cdot {\bf G} &=&{\bar{{\bf H}}}\cdot {\bf H}-{\ \bar{{\bf G%
}}}\cdot {\bf G}  \label{eq:gheq1} \\
(\nabla -{\bf G})\times {\bf G} &=&{\bar{{\bf H}}}\times {\bf H}-{\bar{{\bf G%
}}}\times {\bf G}  \label{eq:gheq2} \\
(\nabla -{\bf G})\cdot {\bf H} &=&%
{\textstyle {1 \over 2}}
({\bf G}-{\bar{{\bf G}}})\cdot {\bf H}  \label{eq:gheq3} \\
\nabla \times {\bf H} &=&-%
{\textstyle {1 \over 2}}
({\bf G}+{\bar{{\bf G}}})\times {\bf H.}  \label{eq:gheq4}
\end{eqnarray}

{\ We introduce a complex triad of basis vectors }$z_o=\ell ,$ $z_{+}=m$ and 
$z_{-}=\bar{m}$. In close analogy with the null tetrad in space-time, the
normalization is chosen 
\[
\ell \cdot \ell =m\cdot \bar{m}=1, 
\]
while all other independent products of the triad vectors vanish\cite
{PerjesJMP}. In their role as linear operators, the triad vectors are
denoted $\left\{ \ell ,m,\bar{m}\right\} =\left\{ D,\delta ,\overline{\delta 
}\right\} $ and have the commutators 
\begin{mathletters}
\label{c}
\begin{eqnarray}
D\delta -\delta D &=&\kappa D+(\overline{\rho }+\epsilon )\delta +\sigma 
\overline{\delta }  \label{c1} \\
\delta \overline{\delta }-\overline{\delta }\delta &=&(\overline{\rho }-\rho
)D+\overline{\tau }\overline{\delta }-\tau \delta .  \label{c2}
\end{eqnarray}
When the triad components of the field equations are taken, there result
five complex Ricci equations from (\ref{eq:gheq0}), of the following form, 
\end{mathletters}
\begin{mathletters}
\label{p}
\begin{eqnarray}
D\sigma -\delta \kappa &=&(\rho +\overline{\rho }+2\epsilon )\sigma +%
\overline{\tau }\kappa +\kappa ^2+2G_{+}\overline{G}_{+}-2\overline{H}%
_{+}H_{+}  \label{pa} \\
D\rho -\overline{\delta }\kappa &=&\rho ^2+\sigma \overline{\sigma }+(%
\overline{\kappa }-\tau )\kappa +G_o\overline{G}_o-H_o\overline{H}_o
\label{pb} \\
D\tau -\overline{\delta }\epsilon &=&(\overline{\kappa }+\tau )\rho +(%
\overline{\kappa }-\tau )\epsilon -(\kappa +\overline{\tau })\overline{%
\sigma }+G_o\overline{G}_{-}+\overline{G}_oG_{-}-H_o\overline{H}_{-}-%
\overline{H}_oH_{-}  \label{pc} \\
\overline{\delta }\sigma -\delta \rho &=&2\tau \sigma +\kappa (\rho -%
\overline{\rho })-\overline{G}_oG_{+}-G_o\overline{G}_{+}+\overline{H}%
_oH_{+}+H_o\overline{H}_{+}  \label{pd} \\
\delta \tau +\overline{\delta }\overline{\tau } &=&\rho \overline{\rho }%
-\sigma \overline{\sigma }+2\tau \overline{\tau }-\epsilon (\rho -\overline{%
\rho })-G_o\overline{G}_o+G_{+}\overline{G}_{-}+G_{-}\overline{G}_{+}+H_o%
\overline{H}_o-\overline{H}_{+}H_{-}-\overline{H}_{-}H_{+},  \label{pe}
\end{eqnarray}
one equation each from (\ref{eq:gheq1}) and (\ref{eq:gheq3}), furthermore
three equations each from (\ref{eq:gheq2}) and (\ref{eq:gheq4}): 
\end{mathletters}
\begin{eqnarray}
&&(D-\rho -\overline{\rho })G_o+(\overline{\delta }+\overline{\kappa }-\tau
)G_{+}+(\delta +\kappa -\overline{\tau })G_{-}  \label{G1} \\
&=&(G_o-\overline{G}_o)G_o+(G_{+}-\overline{G}_{+})G_{-}+(G_{-}-\overline{G}%
_{-})G_{+}+\overline{H}_oH_o+\overline{H}_{+}H_{-}+\overline{H}_{-}H_{+} 
\nonumber
\end{eqnarray}
\begin{mathletters}
\label{G}
\begin{eqnarray}
(\overline{\delta }+\overline{\kappa })G_o-(D-\rho +\epsilon )G_{-}+%
\overline{\sigma }G_{+} &=&\overline{G}_oG_{-}-\overline{G}_{-}G_o-\overline{%
H}_oH_{-}+\overline{H}_{-}H_o  \label{G2} \\
(\delta +\kappa )G_o-(D-\overline{\rho }-\epsilon )G_{+}+\sigma G_{-} &=&%
\overline{G}_oG_{+}-\overline{G}_{+}G_o-\overline{H}_oH_{+}+\overline{H}%
_{+}H_o  \label{G3} \\
(\overline{\delta }-\tau )G_{+}-(\delta -\overline{\tau })G_{-}+(\overline{%
\rho }-\rho )G_o &=&\overline{G}_{+}G_{-}-\overline{G}_{-}G_{+}-\overline{H}%
_{+}H_{-}+\overline{H}_{-}H_{+}  \label{G4}
\end{eqnarray}
\end{mathletters}
\begin{eqnarray}
&&(D-\rho -\overline{\rho })H_o+(\overline{\delta }+\overline{\kappa }-\tau
)H_{+}+(\delta +\kappa -\overline{\tau })H_{-}  \label{H1} \\
&=&%
{\textstyle {1 \over 2}}
\left[ (3G_o-\overline{G}_o)H_o+(3G_{+}-\overline{G}_{+})H_{-}+(3G_{-}-%
\overline{G}_{-})H_{+}\right]  \nonumber
\end{eqnarray}
\begin{mathletters}
\label{Hx}
\begin{eqnarray}
(\overline{\delta }+\overline{\kappa })H_o-(D-\rho +\epsilon )H_{-}+%
\overline{\sigma }H_{+} &=&%
{\textstyle {1 \over 2}}
\left[ (G_o+\overline{G}_o)H_{-}-(G_{-}+\overline{G}_{-})H_o\right]
\label{H2} \\
(\delta +\kappa )H_o-(D-\overline{\rho }-\epsilon )H_{+}+\sigma H_{-} &=&%
{\textstyle {1 \over 2}}
\left[ (G_o+\overline{G}_o)H_{+}-(G_{+}+\overline{G}_{+})H_o\right]
\label{H3} \\
(\overline{\delta }-\tau )H_{+}-(\delta -\overline{\tau })H_{-}+(\overline{%
\rho }-\rho )H_o &=&%
{\textstyle {1 \over 2}}
\left[ (G_{+}+\overline{G}_{+})H_{-}-(G_{-}+\overline{G}_{-})H_{+}\right] .
\label{H4}
\end{eqnarray}

These relations contain the four complex Ricci rotation coefficients $\kappa
=m_{\mu ;\nu }\ell ^\mu \ell ^\nu ,$ $\rho =m_{\mu ;\nu }\ell ^\mu \overline{%
m}^\nu ,$ $\sigma =m_{\mu ;\nu }\ell ^\mu m^\nu $ and $\tau =m_{\mu ;\nu }%
\overline{m}^\mu \overline{m}^\nu ,$ and the imaginary rotation coefficient $%
\epsilon =m_{\mu ;\nu }\overline{m}^\mu \ell ^\nu $ . (Note that the complex
conjugate of, say, $G_{+}$ is $\overline{G}_{-}$).

A solution of particular relevance for us is the Kerr-Newman metric with
mass ${\rm m}$, rotation parameter $a$, electric charge ${\rm e}$ and Ernst
potentials 
\end{mathletters}
\[
{\cal E=}1-%
{\textstyle {2{\rm m} \over \zeta}}
,\qquad \Phi =%
{\textstyle {{\rm e} \over \zeta}}
\]
where 
\[
\zeta =r-ia\cos \vartheta . 
\]
Given the two Killing vectors $\partial /\partial t$ and $\partial /\partial
\varphi $ of the space-time, the three-space may be defined with respect to
any of these two (or their linear combinations). In what follows, we choose
the three-space to be positive-definite.

The orientation of the triad vectors can be chosen at will. Here we adopt a
triad for which the {\it eigenray condition}\cite{PerjesJMP} holds, 
\begin{equation}
G_{+}=0,  \label{G+}
\end{equation}
thereby fixing the direction of the vector $\ell .$ We then have 
\begin{eqnarray}
D &=&%
{\textstyle {\partial  \over \partial r}}
\\
\delta &=&%
{\textstyle {1 \over \left( 2f\right) ^{1/2}\overline{\zeta }}}
\left( -ia%
{\textstyle {\partial  \over \partial r}}
+%
{\textstyle {\partial  \over \partial \vartheta }}
+%
{\textstyle {i \over \sin \vartheta }}
{\textstyle {\partial  \over \partial \varphi }}
\right) .  \nonumber
\end{eqnarray}
Here $f={\rm Re}{\cal E}+{\bar{\Phi}}\Phi .$ For the charged Kerr solution,
the following quantities vanish, 
\begin{equation}
\kappa =\sigma =H_{+}=0.  \label{small}
\end{equation}

\section{The perturbed electrovacuum}

When the perturbed space-time is either axisymmetric or stationary, the
description in Sec. 2 holds once again. In particular, the three-space and
the complex fields ${\bf G}$ and ${\bf H}$ do exist. The perturbed triad is
chosen to satisfy the eigenray condition (\ref{G+}) once again. Three of the
fields, $\kappa ,\sigma $ and $H_{+}$ are small since they vanish for the
unperturbed space-time.

In this gauge, it is a straightforward matter to derive an uncoupled wave
equation. Taking the sum of Eqs. (\ref{H1}) and (\ref{H4}), we get 
\begin{equation}
(D-2\rho )H_o+2(\overline{\delta }-\tau )H_{+}-G_{-}H_{+}+\kappa H_{-}+%
{\textstyle {1 \over 2}}
(\overline{G}_o-3G_o)H_o=0.  \label{cdeHp}
\end{equation}
The commutator (\ref{c1}) is applied on the complex function $H_o$ and the
derivatives $DH_o$ and $\delta H_o$ are eliminated by use of the field
equations (\ref{cdeHp}) and (\ref{H3}), respectively. The derivative $DG_o$
is expressed from the sum of Eqs. (\ref{G1}) and (\ref{G4}), and the
derivatives $DG_{-}$ , $DH_{-}$ , $\delta G_o$ , $\delta H_o$, $\delta G_{-}$
, $\delta H_{-}$, $\delta \rho $ and $D\tau $ from the respective equations (%
\ref{G2}), (\ref{H2}), (\ref{G3}), (\ref{H3}), (\ref{G4}), (\ref{H4}), (\ref
{pd}) and (\ref{pc}). In this way, all purely unperturbed terms can be
removed by the remaining field equations, with the result 
\begin{eqnarray}
&&\left\{ G_o\left[ DD+2\delta \bar{\delta}-D(\bar{\rho}+\epsilon
)+(G_o+2\rho )\epsilon \right] \right.  \nonumber \\
&&-\left[ (G_o-\bar{G}_o+2\rho +2\bar{\rho}+2\epsilon )G_o-2\bar{H}%
_oH_o\right] D  \nonumber \\
&&-(G_{-}+2\tau )G_o\delta +(\bar{G}_{+}G_o+2\bar{H}_{+}H_o)\bar{\delta} \\
&&-G_o\left[ (%
{\textstyle {1 \over 4}}
G_o-\rho )G_o-H_{-}\bar{H}_{+}-(\bar{\rho}+\epsilon )^2+\bar{\tau}%
G_{-}-2\rho \bar{\rho}\right]  \nonumber \\
&&+G_o\left[ (%
{\textstyle {1 \over 2}}
G_{-}-\tau )\bar{G}_{+}-(%
{\textstyle {3 \over 2}}
G_o-%
{\textstyle {3 \over 4}}
\bar{G}_o+\rho +\epsilon )\bar{G}_o-2\delta \tau \right]  \nonumber \\
&&\left. -[2\tau \bar{H}_{+}+3G_{-}\bar{H}_{+}-3\bar{H}_o\bar{G}_o+2(\rho
+\epsilon )\bar{H}_o]H_o\right\} H_{+}  \nonumber \\
&=&\left[ (G_{-}H_o-2G_oH_{-}+2\bar{\delta}H_o)G_o-2H_o\bar{\delta}%
G_o\right] \sigma .  \nonumber
\end{eqnarray}
All terms on the left contain a factor $H_{+}$ and those on the right
contain the rotation coefficient $\sigma .$ Both of these quantities are of
first order, thus the operators acting on them and the factors can be taken
to have their values in the charged Kerr metric. When this is done for the
perturbation function 
\begin{equation}
\phi =H_{+}({\rm e}^2\bar{\zeta}^{-1}+{\rm m})  \label{Phidef}
\end{equation}
the terms on the right-hand side cancel and an uncoupled separable wave
equation results: 
\begin{equation}
\Box _1\phi =0  \label{Phieq}
\end{equation}
where the wave operator is defined 
\begin{eqnarray}
\Box _s &=&\Delta ^{-s}%
{\textstyle {\partial \over \partial r}}
\Delta ^{s+1}%
{\textstyle {\partial \over \partial r}}
+\sin ^{-1}\vartheta 
{\textstyle {\partial \over \partial \vartheta }}
\sin \vartheta 
{\textstyle {\partial \over \partial \vartheta }}
+s\left( s+1-%
{\textstyle {s \over \sin ^2\vartheta }}
\right)  \nonumber \\
&&-2a%
{\textstyle {\partial ^2 \over \partial r\partial \varphi }}
{\cal +}%
{\textstyle {1 \over \sin ^2\vartheta }}
{\textstyle {\partial ^2 \over \partial \varphi ^2}}
+2is%
{\textstyle {\cos \vartheta  \over \sin ^2\vartheta }}
{\textstyle {\partial \over \partial \varphi }}
\end{eqnarray}
and $\Delta =r^2-2{\rm m}r+a^2+{\rm e}^2$ is the horizon function.

Next, applying the commutator (\ref{c1}) on the function $G_o$ , we get the
relation 
\begin{eqnarray}
D\kappa &=&(\epsilon -\rho )\kappa +(4\tau +2G_{-}-2\overline{\delta })\sigma
\nonumber \\
&&-%
{\textstyle {2 \over G_o}}
(\bar{H}_oDH_{+}+\bar{H}_{+}\overline{\delta }H_{+}+\sigma \overline{\delta }%
G_o)  \label{Dka} \\
&&+%
{\textstyle {H_{+} \over G_o}}
[(5G_o-3\bar{G}_o+2\rho +2\epsilon )\bar{H}_o+(3G_{-}+2\tau )\bar{H}_{+}]. 
\nonumber
\end{eqnarray}
Now we act with the commutator (\ref{c1}) on $\kappa $ and eliminate the
derivatives $D\kappa $ and $\delta \kappa $ by use of (\ref{Dka}) and (\ref
{pa}) respectively. The pure first-order relation (\ref{pa}) is one of the
five Ricci equations\cite{PerjesJMP}. We obtain another second-order
differential equation 
\begin{eqnarray}
&&G_o\left\{ G_o\left[ (2G_o-\rho )\rho +4\epsilon ^2+\bar{\rho}^2-2\bar{H}%
_{+}H_{-}+2\bar{G}_{+}G_{-}-4\delta \tau \right] \right.  \nonumber \\
&&+2\delta \bar{\delta}G_o+G_o\left( 2\delta \bar{\delta}+DD+2\bar{\tau}\bar{%
\delta}-D\rho -2D\epsilon -D\bar{\rho}\right)  \nonumber \\
&&\left. +2G_o\left[ (2\epsilon -G_o)\bar{\rho}-2(G_{-}+\tau )\bar{\tau}%
-(G_{-}+2\tau )\delta -(\bar{\rho}+2\epsilon )D\right] \right\} \sigma 
\nonumber \\
&&-2\bar{\delta}G_o\left( \bar{H}_{+}H_o-\bar{\tau}G_o-\bar{G}%
_{+}G_o-G_o\delta \right) \sigma  \nonumber \\
&=&\left\{ G_o\left[ \bar{H}_o\left( 2\delta \epsilon -2\delta D-3\delta 
\bar{G}_o\right) +2\bar{H}_{+}(\delta \tau -\delta \bar{\delta}%
)+(3G_{-}+2\tau )\delta \bar{H}_{+}\right] \right.  \nonumber \\
&&+(3G_oH_{-}-2H_o\tau -3G_{-}H_o)\bar{H}_{+}^2+(3G_o+2\rho +2\epsilon -3%
\bar{G}_o)G_o\delta \bar{H}_o  \nonumber \\
&&-\left[ (3\bar{G}_{+}\bar{H}_o-\bar{H}_{+}G_o)G_o-3(\bar{H}_{+}H_o-\bar{%
\tau}G_o)\bar{H}_o\right] \bar{G}_o  \nonumber \\
&&+\left[ (G_o+2\rho +2\epsilon )\bar{H}_o+2\bar{H}_{+}\tau \right] \bar{G}%
_{+}G_o  \nonumber \\
&&+\left[ (5G_o+2\rho +2\epsilon )\bar{\tau}G_o-2(G_o+\rho +\epsilon )\bar{H}%
_{+}H_o\right] \bar{H}_o  \nonumber \\
&&+\left[ (3G_{-}+2\tau )\bar{H}_{+}-3\bar{G}_o\bar{H}_o+(5G_o+2\rho
+2\epsilon )\bar{H}_o\right] G_o\delta  \nonumber \\
&&+2\left[ (\bar{H}_{+}H_o-\bar{\tau}G_o-\bar{G}_{+}G_o)\bar{H}%
_{+}-G_o\delta \bar{H}_{+}\right] \bar{\delta}  \nonumber \\
&&-2\left[ (\bar{G}_{+}\bar{H}_o+\bar{H}_{+}G_o+\delta \bar{H}_o)G_o-(\bar{H}%
_{+}H_o-\bar{\tau}G_o)\bar{H}_o\right] D  \nonumber \\
&&+\left. \left[ (G_o-5\rho +2\epsilon +3\bar{\rho})G_o+2(3G_{-}+\tau )\bar{%
\tau}\right] \bar{H}_{+}G_o\right\} H_{+}.
\end{eqnarray}

Each term on the left contains the small function 
\begin{equation}
\psi =\sigma G_o/({\rm e}^2\bar{\zeta}^{-1}+{\rm m})  \label{psidef}
\end{equation}
and each term on the right contains an $H_{+}$, to be expressed in terms of $%
\phi .$When inserting the unperturbed values of the operators and factors,
neither the $\psi $ terms, nor the $\phi $ terms cancel, and what we get is
the wave equation 
\begin{equation}
\Box _2\psi =J(\phi ).  \label{psieq}
\end{equation}
Thus the terms containing a solution $\phi $ of Eq. (\ref{Phieq}) will
provide the source function $J(\phi )$ for the equation for $\psi .$ The
source term is a functional of the field $\phi $ containing up to second
derivatives. By considering the differential structure of $J(\phi )$ in the
same gauge, it is possible to derive a decoupled equation also for $\psi ,$
which, however, is quite lengthy.

Given a solution of Eqs. (\ref{Phieq}) and (\ref{psieq}), the perturbation
functions $\sigma $ and $H_{+}$ are available from the simple relations (\ref
{psidef}) and (\ref{Phidef}), respectively. One can next compute the
first-order function $\kappa $ by integrating Eq. (\ref{Dka}). Continuing
the step-by-step integration procedure, we have an algorithm for
systematically getting the full description of the perturbed space-time.

Choosing a different gauge with the roles of the fields $G_{+}$ and $H_{+}$
interchanged, one gets an uncoupled wave equation for the gravitational
perturbation and one with a source term for the electromagnetic perturbation.

\section{Conclusions}

The present description of the simultaneous gravitational and
electromagnetic excitations of a charged black hole already provides a
framework for devising models of astrophysical sources of radiation in
situations where it is sufficient to consider either axisymmetric or
stationary waves. The procedure for general perturbations is more involved
and relies on a gauge adapted to the Killing bivector of the black-hole
background. The details will be given in a follow-up paper.

\section{Acknowledgments}

I thank M\'{a}ty\'{a}s Vas\'{u}th for discussions. This work has been
supported by the OTKA grant T031724.

\end{document}